\documentclass{article}
\usepackage{emulateapj}
\usepackage{apjfonts}

\begin{document}

\submitted{To appear in The Astrophysical Journal}

\title{Measurements of Far-UV Emission from Elliptical Galaxies 
at $z=0.375^1$}

% USE FULL NAME 

\author{Thomas M. Brown$^2$, Henry C. Ferguson$^3$, Jean-Michel 
Deharveng$^4$,\\ and Robert I. Jedrzejewski$^3$}

\begin{abstract}

The ``UV upturn'' is a sharp rise in spectra of elliptical galaxies 
shortward of rest-frame 2500~\.{A}. It is a ubiquitous phenomenon in nearby
giant elliptical galaxies, and is thought to arise primarily from low-mass
evolved stars on the extreme horizontal branch and beyond.
Models suggest that the UV upturn is a very strong function of age for 
these old stellar populations, increasing as the galaxy gets older. 
In some models the change in UV/optical flux ratio is a factor of 25
over timescales of less than 3 Gyr.
To test the predictions for rapid evolution of the UV upturn, we have observed
a sample of normal elliptical galaxies in the $z=0.375$ cluster Abell~370
with the {\it Faint Object Camera} aboard the {\it Hubble Space Telescope}. A
combination of two long-pass filters was used to isolate wavelengths
shortward of rest-frame 2700~\.{A}, providing a measurement of the
UV upturn at a lookback time of approximately 4~Gyr. Surprisingly,
the four elliptical galaxies observed show a range of UV upturn strength
that is similar to that seen in nearby elliptical galaxies, with an
equivalent $m_{1550}-V$ color ranging from 2.9--3.4 mag. Our result
is inconsistent with some models for the UV upturn; other models
are consistent only for a high redshift of formation ($z_f \ge 4$). 

\end{abstract}

\keywords{galaxies: abundances --- galaxies: evolution --- galaxies: stellar
content --- ultraviolet: galaxies --- ultraviolet: stars}

\section{INTRODUCTION}

The spectra of elliptical galaxies and spiral galaxy bulges exhibit a strong
upturn shortward of 2500~\.{A}, dubbed the ``UV upturn.''   The phenomenon was
among the first major discoveries in UV extragalactic astronomy, and the
implied existence of a hot stellar component in elliptical galaxies appeared to
contradict the traditional picture of these galaxies as old, cool,
passively-evolving populations.  The pioneering UV observations -- with 
{\it OAO} (Code \&
Welch 1979) and {\it IUE} (Bertola,
Capaccioli, \& Oke 1982) -- could only sample the Rayleigh-Jeans tail of the
hot UV spectrum ($\lambda > 1200$~\.{A}), with poor signal-to-noise ratios and
resolution. Thus, early explanations for  the UV upturn  included young
massive stars, extreme horizontal branch (EHB) stars, post-asymptotic giant
branch (PAGB) stars and
several binary scenarios (Greggio \& Renzini 1990). 
Eventually, it became clear that the phenomenon was likely due to the
presence of hot {\it evolved} stars (see Greggio \& Renzini 1990; Bressan,
Chiosi, \& Fagotto 1994; Horch, Demarque, \& Pinsonneault 1992), and thus the
UV upturn is still consistent with the picture of elliptical galaxies 
as passively
evolving populations.  Observational evidence for this view was found
in  the spectra from the {\it Hopkins Ultraviolet Telescope (HUT)}
(Ferguson et al.\ 1991; Brown et  al.\
1997); these spectra are reproduced
well by a composite  population of EHB, post-EHB, and PAGB  stars.

\vskip 3em

{\small
$^1$Based on observations with the NASA/ESA Hubble Space
Telescope obtained at the Space Telescope Science Institute, which is operated
by the Association of Universities for Research in Astronomy, Incorporated,
under NASA contract NAS~5-26555.

$^2$Laboratory for Astronomy \& Solar Physics, Code 681, NASA/GSFC,
Greenbelt, MD 20771. tbrown@pulsar.gsfc.nasa.gov.
 
$^3$Space Telescope Science Institute, 3700 San
Martin Drive, Baltimore, MD 21218. ferguson@stsci.edu \& rij@stsci.edu.
 
$^4$Laboratoire d'Astronomie Spatiale du
CNRS, Traverse du Siphon, BP 8, F-13376 Marseille Cedex 12, France.
jmd@astrsp-mrs.fr.
}

Characterized by the $m_{1550}-V$ color, the UV upturn varies strongly
(ranging over 2.05--4.50 mag) 
in  nearby quiescent early-type galaxies (Bertola et  al.\
1982; Burstein et al.\ 1988), even though the spectra of elliptical galaxies  
at longer  wavelengths are qualitatively very similar.  
The $m_{1550}-V$ color is positively correlated with
the strength of Mg$_2$ line absorption (i.e.\ bluer at higher line strengths),
opposite to the behavior of optical color indices; it also correlates with
velocity dispersion and luminosity, but to a weaker extent (Burstein et al.\
1988).

The UV upturn is thought to ``turn on'' when the stars in an elliptical 
are of sufficient age to populate the extreme horizontal branch 
(see Greggio \& Renzini 1990; Bressan et al.\ 1994). After leaving the EHB, 
the post-EHB stars  become hotter and brighter before descending the white 
dwarf cooling curve (see Dorman, Rood, \& O'Connell 1993).
EHB stars and their associated post-EHB phases are
considerably longer lived than the bright PAGB stars originating on the red
end of the horizontal branch (HB). 
Thus, they can dominate the spectrum of an elliptical galaxy
despite their minority status in the stellar population.  The varying fraction
of EHB stars in the nearby elliptical galaxy
populations may drive the variation in
the  strength of the UV upturn (see Brown et al.\ 1997 and references
therein).

Because the main sequence turnoff 
mass and HB morphology are sensitive to the
age of a passively evolving population, evolution theory
predicts that the UV upturn should fade with lookback
time, with the amount of fading depending upon the mechanisms driving HB
morphology and the redshift of formation.  The UV upturn may be the
most rapidly evolving feature in elliptical galaxies (see Greggio \& Renzini
1990).  Several previous attempts to measure the 
UV upturn at intermediate redshifts, using UV imaging and spectroscopy with 
the {\it Hubble Space Telescope (HST)}, have been inconclusive (e.g., Windhorst
et al.\ 1994; Renzini 1996; Buson et al.\ 1998).  
We have observed the cluster Abell~370 (at $z=0.375$) with
the hope of measuring the UV upturn at an epoch considerably younger than our
own; the lookback time for this redshift is 4 Gyr (we
assume a cosmology of $H_o = 67$~km~s$^{-1}$, $q_o = 0.05$, and $\Lambda = 0$
throughout this Letter).  We used a combination of long-pass filters
in the {\it Faint Object Camera (FOC)} aboard 
{\it HST} to create a synthetic bandpass with high UV throughput.
Our four fields were each
centered on one or two giant elliptical galaxies, although fainter companions
are also included in the images. The galaxies under study are all 
cluster members, as confirmed
via ground-based spectroscopy, and show no spectroscopic
indications of star formation (Soucail et al.\ 1988).
Their $U-V$ colors are consistent with passive evolution (MacLaren, Ellis,
\& Couch 1988).  Morphological classification is confirmed by
{\it Wide Field and Planetary Camera 1 (WFPC1)} 
(Couch et al.\ 1994) and {\it WFPC2} (GO proposal 6003) imaging.

\noindent
\begin{center} Table 1: {\it FOC} Observations \end{center}
%\begin{table}
%\caption{{\it FOC} Observations}
%\label{tabobs}
\begin{tabular}{|l||r|r|r|r|}
\tableline
 & R.A. & Dec. & \multicolumn{2}{c|}{Exposure (sec)} \\
Field & (J2000) & (J2000) & F130LP & F370LP \\ 
\tableline \tableline
A370-1 & $2^h39^m53^s$ & $-1^\circ 34\arcmin 54\arcsec$ & 5411 & 2794\\
A370-2 & $2^h39^m53^s$ & $-1^\circ 34\arcmin 17\arcsec$ & 5314 & 5649\\
A370-3 & $2^h34^m52^s$ & $-1^\circ 33\arcmin 42\arcsec$ & 5314 & 8502\\
A370-4 & $2^h39^m51^s$ & $-1^\circ 33\arcmin 52\arcsec$ & 5314 & 8502\\
\tableline
G158-100 & $0^h33^m55^s$ & $-12^\circ 07\arcmin 57\arcsec$ & 1077 & 1162\\
CR-831 & $16^h24^m10^s$ & $-26^\circ 33\arcmin 12\arcsec$ & 1077 & 1186 \\
\tableline
\end{tabular}
%\end{table}

\section{OBSERVATIONS} \label{secobs}

Over the course of 1997, 
we observed four fields in Abell~370, plus two calibration stars (Table 1).  
The cluster fields were chosen to place at least
one bright giant elliptical in the center of each image.
Fields 1, 2, 3, and 4 are
centered on galaxies BOW 9, 10, 24, and 34, respectively (Butcher,
Oemler, \& Wells 1983).  BOW 24 is actually two elliptical galaxies, and so
we refer to them as 24a and 24b; 24b is $2\arcsec$ directly south of 24a.
The fields suffer from little foreground extinction:
$E(B-V) \le 0.03$~mag (Burstein \& Heiles 1984; Schlegel, Finkbeiner, 
\& Davis 1998).

The {\it FOC} is a photon-counting imager with a high sensitivity at the short
end of the {\it HST} wavelength range.  Our use of 
the $512 \times 1024$ zoomed format provided  the full $14 \times
14\arcsec$ field of view at the expense of the full dynamic range available in
the {\it FOC}.  The F130LP and F370LP filters are designed to block light
shortward of 1300~\.{A} and 3700~\.{A}, respectively. Longward of the cutoff
wavelength, the transmittance of each filter rapidly rises to a value that is
nearly wavelength independent (0.92 in F130LP and 0.83 in F370LP);
the ratio of the light detected in each filter is extremely sensitive to the
flux shortward of 3700~\.{A} and insensitive to the shape of a spectrum
longward of 3700~\.{A}. This wavelength corresponds to
2690~\.{A} in the rest frame of Abell~370, and thus these filters are useful 
for measuring the UV upturn at this redshift.

All of our images were processed via the standard pipeline, which includes
dezooming, geometric correction, and flat-fielding. {\it FOC} images also
contain reseau marks where the
flux is suppressed but not corrected in the pipeline.  
Before registering and co-adding our images,
we corrected the reseau marks by
interpolating from neighboring unaffected pixels.
The affect of our reseau correction is very small,
because the individual frames were shifted by
several pixels, and because no reseau marks fell near the core of 
any galaxy of interest.

Unfortunately, one of the bright companion galaxies, BOW 58,  fell behind
one of the {\it FOC} occulting ``fingers'' in the image containing BOW 10.
Also, our {\it FOC} images demonstrate that
BOW 9, the central galaxy of Frame 1, is a close superposition
of two galaxies -- a giant elliptical and a small bright companion.  Because
the small companion is so close to the  giant elliptical, we cannot
disentangle their UV light, and so we exclude BOW 9 from our analysis at
this time.

\section{PHOTOMETRY} \label{secphot}

We used the IRAF routine PHOT to perform aperture photometry on the bright
elliptical galaxies 
in our images.  The point spread functions of the two filters
are similar, and so taking the ratio of the F130LP countrate to the  F370LP
countrate requires no aperture correction, given a suitably large  aperture;
our aperture has a radius of 65 pixels (0.91$\arcsec$).   The
background annulus was chosen to avoid neighboring  galaxies.
The annulus inner and outer radii were 165 and 195 for BOW 10;
for the other galaxies the radii were 145 and 165.  

Our large apertures enclose practically all of the detectable UV light in these
galaxies.  Sixty-five {\it FOC} pixels correspond
to 4.55~kpc at $z=0.375$. Compared to the {\it IUE} and {\it HUT} apertures
used to measure the UV  light in nearby elliptical galaxies, our apertures are
quite large.
In their survey of the UV upturn in nearby galaxies, Burstein et al.\
(1988) matched the $10 \times 20\arcsec$ {\it IUE} aperture with a $7\arcsec$
radius aperture in the optical; $7\arcsec$ subtends 0.85 kpc at the distance
of NGC~1399, and so the equivalent aperture radius at $z=0.375$ would be
12 {\it FOC} pixels. Although we note the UV light extends significantly
beyond 12 {\it FOC} pixels, for comparison to nearby elliptical galaxies, 
we also performed our photometry with this smaller aperture (but with no change
in the sky annulus); this gives an
indication of the nuclear UV upturn strength in the Abell~370 elliptical 
galaxies.
Table~2 gives the {\it FOC} count rates in both the large and small
apertures.  Statistical errors are all on the order of 1\%.

\noindent
\begin{center} Table 2: {\it FOC} Photometry and Model Predictions \end{center}
%\begin{table}
%\caption{{\it FOC} Photometry and Model Predictions}
%\label{tabphot}
\begin{tabular}{|l||r|r|r|r|}
\tableline
          & F130LP        & F370LP        & F130LP/ & $m_{1550}-V$  \\
Galaxy & cts s$^{-1} $ & cts s$^{-1} $ & F370LP & (mag) \\ 
\tableline \tableline
\multicolumn{5}{|c|}{Photometry: 4.55 kpc aperture radius} \\
\tableline
BOW 10 & 9.03 & 6.82 & 1.32 & 3.0 \\
BOW 24a & 4.55 & 3.69 & 1.23 & 3.4 \\
BOW 24b & 4.22 & 3.14 & 1.34 & 2.9 \\
BOW 34a & 7.13 & 5.75 & 1.24 & 3.4 \\
\tableline
\multicolumn{5}{|c|}{Photometry: 0.85 kpc aperture radius}\\
\tableline
BOW 10 & 1.78 & 1.21 & 1.47 & 2.2 \\
BOW 24a & 0.841 & 0.650 & 1.29 & 3.1 \\
BOW 24b & 1.11 & 0.840 & 1.32 & 2.9 \\
BOW 34a & 1.48 & 1.16 & 1.28 & 3.2 \\
\tableline
\multicolumn{5}{|c|}{Non-evolving templates: 0.85 kpc aperture radius}\\
\tableline
NGC 1399 & 1.55 & 1.02 & 1.52 & 2.05 \\
M 60 & 1.26 & 0.856 & 1.47 & 2.24 \\
M 49 & 1.22 & 0.995 & 1.23 & 3.42 \\
M 49 (no UV) & 1.10 & 0.995 & 1.11 & $\infty$ \\
\tableline
\end{tabular}
%\end{table}

\smallskip

Photometry on the star G158-100 demonstrates that the
{\it FOC} throughputs are in agreement with expectations. We extended
the optical (3200--9200~\.{A})  spectrum of this white dwarf (Colina \& Bohlin
1994; Oke 1990) into the UV using a Kurucz (1992) synthetic spectrum with
parameters  T$_{\rm eff}$~=~5250~K, log~$g=5$ (the maximum available
in the Kurucz grid), and [Me/H]$=-3.0$.  Laird et
al.\ (1988) quote a a somewhat cooler temperature of 5072~K and a metallicity
of $-2.82$, but we found that our choice of parameters gave the best  agreement
with the optical spectrum.  Note that 
the surface gravity in the model has a negligible effect for
our purposes here.  Using the IRAF/STSDAS routine CALCPHOT in the SYNPHOT
package, we calculated a F130LP/F370LP count rate ratio of 1.691; aperture
photometry on the star yields a count rate ratio of 1.728 (a 2\%
difference).  The photometry aperture radius was 20 pixels,
with a background annulus spanning 60--80 pixels in radius.

CR-831, a UV-bright star in the globular cluster M4
(Drukier, Fahlman, \& Richer 1989), was observed with our
filters in conjunction with the {\it FOC} prism.  These data confirm that
the F130LP/F370LP count rate ratio is essentially flat longward of 3700~\.{A}.
Both
bandpasses are truncated at $\sim$6000~\.{A} by the detector  sensitivity.
Unfortunately, the blue end of each
spectrum falls off the edge of the {\it FOC} detector, and thus the data are
not useful as another check on the absolute calibration.

\section{COMPARISON TO LOCAL ELLIPTICAL GALAXIES}

Translating our F130LP/F370LP ratio to $m_{1550}-V$ cannot be done
in a model-independent manner, so we compare our measurements to non-evolving
templates of elliptical galaxies, and then discuss the effects of evolution.
To create our templates, we folded the spectra of three nearby well-studied
elliptical galaxies (M~49, M~60, and NGC~1399) 
through the {\it FOC} instrument response function, using the
IRAF/STSDAS routine CALCPHOT.  The galaxies
were observed by {\it HUT} in the far-UV (900--1840~\.{A}), by {\it
IUE} in the near-UV (1240--3200~\.{A}), and by the 4000~\.{A} break project
(Kimble, Davidsen, \& Sandage 1989) in the optical (3200--6200~\.{A}). 
We spliced the {\it IUE} and {\it HUT} spectra to
create UV spectra extending across the entire UV range, keeping the {\it HUT}
data in the region of overlap.  The optical data were then extended to
9900~\.{A} by normalizing the redshifted elliptical template of Kinney et al.\
(1996) to the data in the 5000--6200~\.{A} region.  Because the optical
and UV data do not overlap, we spliced them together at 3200~\.{A} after
normalizing each to the flux levels given by Burstein et al.\ (1988).  These
templates thus reproduce the flux and $m_{1550}-V$ in the cores of these
galaxies, as viewed from distances of 21.9, 21.9, and 24 Mpc for M~49,
M~60, and NGC~1399. We then redshifted our templates to $z=0.375$,
applying no corrections for evolution.
The three spectra serve as non-evolving templates for comparison to
Abell~370.

The predicted {\it FOC} count rates for our templates are shown in
Table~2; we use them to construct a relation between $m_{1550}-V$
and F130LP/F370LP, and then interpolate
the approximate $m_{1550}-V$ color for each of our
Abell~370 elliptical galaxies.  
These elliptical galaxies appear at moderate UV
upturn strength; in the large 65-pixel aperture, two are similar to M~49,
and two fall between M~49 and M~60.  Note that F130LP/F370LP would be
1.11 if M~49 produced no flux shortward of rest-frame 2500~\.{A}, and
so all of our Abell~370 measurements are significant detections of UV emission.
Uncertainties from counting statistics are less than 1\%.  From tests
measuring the galaxy fluxes in the individual frames,
with the background level artificially varied by $\pm$1\%,
we estimate that the systematic errors 
can be no more than about 10\%.
With the exception of BOW 24b, the Abell~370 elliptical galaxies
are significantly bluer in $m_{1550}-V$ through the  smaller aperture.
Most nearby elliptical galaxies 
(including M~49, M~60, and NGC~1399) also become
bluer in $m_{1550}-V$ at smaller radii (Ohl et al.\ 1998), hence
the relevance of comparison between equivalent apertures.  

Some flux from the main sequence turnoff is present
shortward of rest-frame 2700~\AA.  Main sequence turnoff 
evolution thus has some effect
on the F130LP/F370LP ratio, but the effect is small for a lookback time
of 4~Gyr.  The massive elliptical
galaxy model of Tantalo et al.\ (1996) serves as an
example.   Without UV upturn stars, a 5 Gyr population 
has F130LP/F370LP = 1.17, and a 10 Gyr population 
has F130LP/F370LP = 1.12 (i.e., a 4\% effect); with a UV upturn present,
the effect is even smaller.

\section{DISCUSSION}

Because they are composed of old, passively evolving populations, elliptical
galaxies offer great promise for tracing the evolution of the Universe.  One
of the major goals in extragalactic research is the determination of the
``redshift of formation,'' $z_f$, that marks the age where most of the stars
in early-type galaxies formed.  
The formation redshift is a parameter that may have
less meaning in a Universe where elliptical
galaxies are formed through hierarchical
merging, but recent studies of galaxy clusters at optical and infrared
wavelengths, out to $z \sim 1$ and  
including Abell 370, are consistent with the monolithic collapse of massive 
cluster galaxies at high redshift
($z > 2-4$), followed by quiescent evolution thereafter (Stanford,
Eisenhardt, \& Dickinson 1998; Kodama et al.\ 1998).  
These studies also demonstrate that 
early-type galaxies in cluster cores show no evidence of recent star 
formation, although there may be some indication for
recent star formation in the galaxies lying on the cluster outskirts 
(Stanford et al.\ 1998 and references therein).

The UV upturn provides a sensitive tracer of age in old
populations, and can potentially constrain $z_f$ with an independent
diagnostic, tracing the hotter populations at shorter wavelengths.
Our observations of elliptical galaxies in Abell~370 show 
no evidence for evolution to a lookback time of approximately
4 Gyr. With only four galaxies at one redshift, it is 
premature to make detailed comparisons to models.
Nevertheless, it is interesting to consider the implications
of the Abell~370 measurements for the redshift of formation of
cluster elliptical galaxies (if the models are valid), and 
to compare the detections qualitatively to the 
expectations from the models.

Figure~1 shows the evolution of the UV upturn as measured by
our F130LP/F370LP ratio (assuming the bandpasses are blue-shifted with
respect to the model spectra by 1.375).
In our assumed cosmology, the current 
age of the Universe is 13 Gyr, and the Abell~370 elliptical
galaxies are observed
in a Universe of age 9~Gyr.   We assume three formation
redshifts: $z_f = 2, 4,$ and 8.
In this example, we again use the infall models
of Tantalo et al.\ (1996).  The Tantalo et al.\ (1996) models
assume that galaxies evolve with infall of primordial gas, and include
the effects of galactic winds.  In their most massive models
(1--3$\times 10^{12} M_{\odot}$), the UV
upturn ``turns on'' at a galactic age of 6 Gyr (with the ascension of 
metal-rich EHB and post-EHB stars), reaches the strength
seen in local galaxies by an age of 7 Gyr, and levels off thereafter
(see Tantalo et al.\ 1996, Fig.~16).

Given that the F130LP/F370LP ratio ranges from 1.23--1.34 in these four
Abell 370 elliptical galaxies, we can infer that they
formed at a redshift of at least 4, assuming the model assumptions are valid.  
Models of lower mass (0.1--5$\times 10^{11} M_{\odot}$) 
do not reach our observed range of UV upturn strength 
before 15 Gyr, and thus are not included in Figure 1.  
Measurements of the 
UV upturn for less luminous elliptical galaxies at this redshift would be 
of great interest.  If the Abell 370 elliptical galaxies and
the local population of elliptical galaxies both formed at a common redshift 
$z \ge 4$, the lack of evolution seen in the UV upturn can be understood
as evidence that both epochs are on the ``flat'' portion of the UV
upturn evolutionary curves,
after the UV upturn has leveled off.

More recent theoretical examples of the UV upturn evolution are 
apparently in disagreement with our Abell~370 observations.  For example,
Tantalo et al.\ (1998) revised their 
earlier models to include gradients in mass density and star formation.  
These models were meant as a tool for exploring galaxy evolution
beyond the simplifying assumptions of a one-zone scheme, and 
several parameters could be tuned to give agreement with our results,
such as the time of onset for galactic winds, the efficiency of the 
star formation rate, and the accretion time scale. 
The main discrepancy between our observations and the integrated colors in
these more recent models is
that in the most massive (and rapidly evolving) models, the rise of the
UV upturn is delayed until 10~Gyr.  The elliptical
galaxies in Abell~370,
observed in a Universe of age 9~Gyr, would not have the time to achieve
their measured UV colors even if they were formed at very large
$z_f$.

A different discrepancy occurs when we compare our data to the theoretical
predictions of Yi et al.\ (1998).  In their models, the UV upturn can
appear at ages as early as 5~Gyr and as late as 10~Gyr, but the $m_{1550}-V$
color increases continuously from 6 to 0~mag.
Today's local elliptical galaxies thus lie along a steep slope in the Yi 
et al.\ (1998) scheme, and should fade very rapidly with lookback time under 
all 5 sets of their models.
Our results are thus strongly inconsistent with the Yi et al.\ (1998)
predictions.

One way around the rather surprising result that the UV upturn
exists at $z=0.375$ is to imagine that the UV emission we are 
seeing comes from star formation. The galaxies selected for 
our study are among the best studied at this redshift and show
no morphological, spectroscopic, or optical photometric evidence
of star formation.  Nevertheless, a star formation rate of approximately only 
0.02 $M_{\odot}$ yr$^{-1}$ would be enough to produce the UV upturn we 
observe (see Madau, Pozzetti, \& Dickinson 1998).  
In this case the hypothesis would be that elliptical galaxies
are still forming stars at a low level at $z = 0.375$, but must
cease to do so by $z=0$. UV observations of clusters at intermediate
redshifts will test whether this is correct.
	 
Our measurements provide a first step in mapping the evolution
of the UV upturn with lookback time. Further observations are
clearly needed to rule out star formation as the source of the
UV emission in these high redshift galaxies, and to trace the 
evolution to both higher and lower redshifts. HST Observations of
a cluster at $z = 0.55$ are planned in the near future (GTO proposal 8020). 
Unless the redshift of formation is very high, these galaxies ought to be 
very faint in the far-UV.

\acknowledgments
Support for this work was provided by NASA through grant number GO-6667 from
the Space Telescope Science Institute, which is operated by the Association of
Universities for Research in Astronomy, Incorporated, under NASA contract
NAS~5-26555. TMB acknowledges support at Goddard Space Flight Center by
NAS~5-6499D.

\parbox{3.25in}{\epsfbox{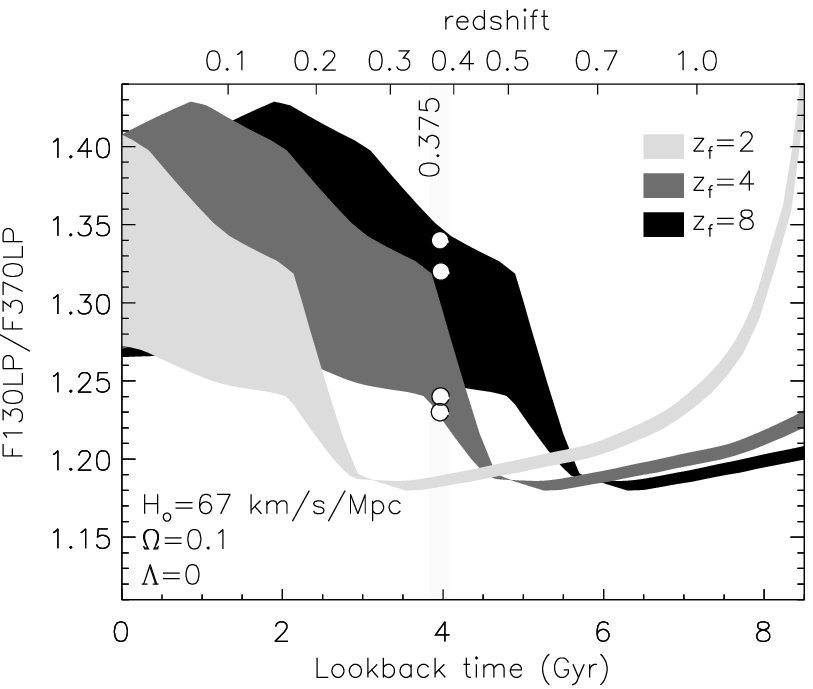}}
\smallskip

\centerline{\parbox{3.25in}{\small {\sc Fig. 1}
The evolution of the F130LP/F370LP flux ratio
(assumed to be blue-shifted by 1.375 with respect to the spectra)
according to the giant elliptical galaxy models 
of Tantalo et al.\ (1996), viewed as a function
of lookback time, assuming a reasonable set of cosmological parameters
and 3 different epochs of galaxy formation (labeled).
The spread in F130LP/F370LP is bounded by models at
$M=3\times 10^{12} M_{\odot}$ and
$M=1\times 10^{12} M_{\odot}$.  Note the sudden onset of UV emission caused by
the appearance of metal-rich EHB stars.  
Our measurements (circles) may indicate $z_f \ge 4$ in Abell~370 
if these models are valid. 
}}

\end{document}